\documentclass[11pt]{article}
\usepackage{amsmath}
\begin{document}

\thispagestyle{empty}

\begin{flushright} LPTENS-05/17 \end{flushright}

\vskip 0.5cm

\begin{center}{\LARGE { GAUGE THEORIES AND  
\vskip 0.4cm
 NON-COMMUTATIVE GEOMETRY}}

\end{center}

\vskip1cm
\begin{center}
{\bf{J. Iliopoulos}}

\vskip0.2cm

Laboratoire de Physique Th\'eorique CNRS-ENS\\ 
24 rue Lhomond, F-75231 Paris Cedex 05, France\\
ilio@lpt.ens.fr

\end{center}

\vskip1.0cm

\begin{center}

{\bf ABSTRACT}

\end{center}
\vskip 0.5cm

It is shown that a $d$-dimensional classical $SU(N)$ Yang-Mills theory can be formulated in a
$d+2$-dimensional space, with the extra two dimensions forming a surface with
non-commutative geometry.

\vskip 5cm

Talk given at the Symposium in Honor of the 70th birthday of Julius Wess, 10-11 January 2005.

Max-Planck Institut f\"{u}r Physik (Werner Heisenberg-Institut), M\"{u}nchen.

\newpage

\section{Introduction}

With Quantum Mechanics, W. Heisenberg introduced non-commutativity in phase space. This property was found to be crucial to solve the problem of the instability of the classical Rutherford atom due to the short-distance singularity of the Coulomb potential. So, when few years later it was realised that these short-distance singularities reappear if relativistic corrections are consistently taken into account, it is not surprising that it was Heisenberg again who first proposed a non-commutative geometry for space itself as a means to provide an effective ultraviolet cut-off \cite{wess1}, \cite{snyder}. As it turned out, this motivation did not prove fruitful for elementary particle physics. With the development  of the renormalisation programme in the framework of quantum field theories, the problem of ultraviolet divergences took a completely different turn. While a space cut-off makes all theories finite, the renormalisation programme applies to very few and very specific field theories. It is a most remarkable fact that they are precisely the ones chosen by Nature. It is not finiteness but rather lack of sensitivity to unknown physics at very short distances that turned out to be the important criterion. The geometry of physical space may still produce an ultraviolet cut-off, but its presence is not relevant for the calculation of physical processes among elementary particles.

However, almost at the same time, a new motivation for studying theories in a non-commutative space appeared, although only recently it was fully appreciated. In 1930 L.D. Landau \cite{landau} solved the problem of the motion of an electron in an external constant magnetic field and, besides computing the energy levels, the so-called ``Landau levels'', he showed that the components of the velocity operator of the electron do not commute. A simple way to visualise this result is to think of the classical case where the electron follows a spiral trajectory whose projection on a plane perpendicular to the field is a circle.  In Landau's quantum mechanical solution its coordinates are:

\begin{equation}
\label{circlecoord}  
x_c=\frac{cp_y}{eH}+x~~~~;~~~~y_c=-\frac{cp_x}{eH}
\end{equation}
which shows that the two coordinates do not commute. The magnetic field has induced a non-commutative structure on space itself. Following Heisenberg's suggestion, R. Peierls \cite{peierls} showed that, at least  the lowest Landau level, can be obtained by using this space non-commutativity.  Since the presence of non-vanishing magnetic-type external fields is a common feature in many modern supergravity and string models, the study of field theories formulated on spaces with non-commutative geometry \cite{connes} has become quite fashionable. A new element was added a few years ago with the work of N. Seiberg and E. Witten \cite{sw} who showed the existence of a map between  gauge theories formulated in  spaces with commuting and non-commuting coordinates. A very incomplete list of recent articles includes \cite{genrev}, \cite{special1}.   Here I want to mention a different but related result which was obtained in collaboration with E.G. Floratos \cite{finew}. I shall state and, to a certain extend, prove, the following statement:

\vskip0.5cm

{\it Statement:} Given an $SU(N)$ Yang-Mills theory  in a $d-$dimensional space
with potentials

\begin{equation}
\label{gaugepot}
 A_{\mu}(x)~=~A_{\mu}^a (x)~t_a 
\end{equation}
where $t_a$ are the standard $SU(N)$ matrices, there exists a reformulation in which 
the gauge fields and the gauge potentials become: 

\begin{equation}
\label{newlimits}
A_{\mu}(x) \rightarrow {\cal A}_{\mu}(x,z_1,z_2)~~~~~~~ F_{\mu \nu}(x) \rightarrow {\cal F}_{\mu \nu}(x,z_1,z_2)
\end{equation}
where ${\cal A}$ and ${\cal F}$ are fields in a $(d+2)-$dimensional space, greek indices still run from 0 to $d-1$ and $z_1$ and $z_2$ are
local coordinates on a  two-dimensional surface endowed with non-commutative
geometry. They will be shown to satisfy the commutation relation

\begin{equation}
\label{fuzsphcom}
[z_1,z_2]=\frac{2i}{N}
\end{equation}

The commutators in  the original $SU(N)$ Yang-Mills theory are
replaced by the Moyal brackets \cite{moyal}, \cite{fairlie} with respect to the non-commuting coordinates.

\begin{equation}
\begin{split}
\label{commoy}
[A_{\mu}(x), A_{\nu}(x)] & \rightarrow \{{\cal A}_{\mu}(x,z_1,z_2), {\cal
    A}_{\nu}(x,z_1,z_2)\}_{Moyal}\\
[A_{\mu}(x),\Omega(x)] & \rightarrow \{{\cal A}_{\mu}(x,z_1,z_2), 
 {\it   \Omega} (x,z_1,z_2)\}_{Moyal}
\end{split}
\end{equation}
where $\Omega $ is the function of the gauge transformation and $\{,\}_{Moyal}$
denotes the Moyal bracket with respect to the two operators $z_1$ and $z_2$.
The trace over the group indices in the original Yang-Mills action becomes a
two dimensional integral over the surface:

\begin{equation} 
\label{action}
\int d^4x~ Tr\left( F_{\mu \nu}(x)F^{\mu \nu}(x)\right)~~\rightarrow~~ \int d^4xdz_1dz_2~{\cal F}_{\mu \nu}(x,z_1,z_2)*{\cal F}^{\mu \nu}(x,z_1,z_2)
\end{equation}

The *-product will be defined later. When $N$ goes to infinity, the two $z$'s
commute and the *-product reduces to the ordinary product.

\vskip 0.5 cm 

In what follows I shall give a
partial proof of this statement.
\vskip 0.5 cm

Let me start by recalling a well-known algebraic result: The Lie algebra of
the group $SU(N)$, at the limit when $N$ goes to infinity, with the generators appropriately 
rescaled, becomes the algebra of the area preserving diffeomorphisms of a
surface. There exist explicit proofs of this theorem for the case of the
sphere and the torus \cite{hoppe}, \cite{fi1}, \cite{zachos1}. This implies a
corresponding field theoretic result: A classical $SU(N)$ Yang-Mills theory on
a $d$-dimensional space at an appropriate large $N$ limit is equivalent to a
field theory on  $d+2$ dimensions with the matrix commutators replaced by
Poisson brackets with respect to the two new coordinates \cite{fi2},
\cite{zachos2}, \cite{sakita}. This is just the large $N$ limit of the
relations (\ref{gaugepot})-(\ref{action}) we want to prove, in other words we
want to establish the equivalence between Yang-Mills
theories and field theories on surfaces to any order in
$1/N$. 

\vskip 0.5 cm

In order to be specific, let us consider the case of the sphere. One way to
prove the algebraic result \cite{fi1}, is to  start with the remark that the spherical harmonics $Y_{l,m}(\theta ,\phi)$ are
harmonic homogeneous polynomials of degree $l$ in three euclidean coordinates $x_{1}$, 
$x_{2}$, $x_{3}$:

\begin{equation}
\label{sphcoord}
x_{1}=cos\phi ~sin\theta,~~~~x_{2}=sin\phi ~sin\theta, ~~~~x_{3}=cos\theta
\end{equation}

\begin{equation}
\label{Ylm}
Y_{l,m} (\theta, \phi)=~~\sum _{i_{k}=1,2,3 \atop k=1,...,l}
\alpha_{i_{1}...i_{l}}^{(m)}~x_{i_{1}}...x_{i_{l}}
\end{equation}
where $\alpha_{i_{1}...i_{l}}^{(m)}$ is a symmetric and traceless tensor. For fixed $l$ there are 
$2l+1$ linearly independent tensors $\alpha_{i_{1}...i_{l}}^{(m)}$,
$m=-l,...,l$. 

Let us now choose, inside $SU(N)$, an $SU(2)$ subgroup by choosing three $N\times N$ hermitian 
matrices which form an $N-$dimensional irreducible representation of the Lie algebra of $SU(2)$.

\begin{equation}
\label{su2}
[S_{i},S_{j}]=i\epsilon_{ijk}S_{k}
\end{equation}

The $S$ matrices, together with the $\alpha$ tensors introduced before, can be
used to construct a basis of $N^2-1$ matrices acting on the fundamental
representation of $SU(N)$ \cite{schwinger}. 

\begin{equation}
\begin{split}
S^{(N)}_{l,m}=~~\sum _{i_{k}=1,2,3 \atop k=1,...,l}
\alpha_{i_{1}...i_{l}}^{(m)}~S_{i_{1}}...S_{i_{l}}  \\
[S^{(N)}_{l,m},~S^{(N)}_{l',m'}]=if^{(N)l'',m''}_{l,m ;~ l',m'}~S^{(N)}_{l'',m''}
\label{sunalg}
\end{split}
\end{equation}
where the $f'$s appearing in the r.h.s. of (\ref {sunalg}) are just the $SU(N)$ structure constants in a 
somehow unusual notation. 
The important, although trivial, observation is that the three $SU(2)$ generators $S_{i}$, 
rescaled by a factor proportional to $1/N$, will have well-defined limits as
$N$ goes to infinity.

\begin{equation}
\label{rescsu2gen}
S_{i}\rightarrow T_{i} = \frac {2}{N} S_{i}
\end{equation}

Indeed, all matrix elements of $T_i$ are bounded by $|(T_i)_{ab}|\leq$ 1. 
They satisfy the rescaled algebra:

\begin{equation}
\label{rescsu2}
[T_{i},T_{j}]=\frac {2i}{N} \epsilon _{ijk}T_{k}
\end{equation}
and the Casimir element

\begin{equation}
\label{casimir}
T^2=T_{1}^2+T_{2}^2+T_{3}^2=1-\frac {1}{N^2}
\end{equation}
in other words,  under the norm $\| x \| ^2 = Trx^2 $, the limits as $N$ goes to infinity of the 
generators $T_{i}$ are three objects $x_{i}$ which commute by (\ref{rescsu2}) and are constrained 
by (\ref{casimir}):

\begin{equation}
\label{constr}
x_{1}^2+x_{2}^2+x_{3}^2=1
\end{equation}

If we consider two polynomial functions $f(x_{1},x_{2},x_{3})$ and $g(x_{1},x_{2},x_{3})$ the corresponding matrix polynomials  $f(T_{1},T_{2},T_{3})$ and $g(T_{1},T_{2},T_{3})$ have commutation relations for large $N$ which follow from (\ref{rescsu2}):

\begin{equation}
\label{limpois}
\frac {N}{2i}~ [f,g] \rightarrow ~~\epsilon_{ijk} ~x_{i}~ \frac {\partial {f}}{\partial {x_{j}}} \frac {\partial {g}}{\partial {x_{k}}} 
\end{equation}

If we replace now in the $SU(N)$ basis (\ref {sunalg}) the $SU(2)$ generators $S_{i}$ by the rescaled ones $T_{i}$, we obtain a set of $N^2-1$ matrices $T^{(N)}_{l,m}$ which, according to (\ref{Ylm}), (\ref{sunalg}) and (\ref{limpois}), satisfy:

\begin{equation}
\label{limpoisY}
\frac {N}{2i}~ [T^{(N)}_{l,m},T^{(N)}_{l',m'}] \rightarrow ~\{ Y_{l,m},Y_{l',m'} \}
\end{equation}

The relation (\ref{limpoisY}) completes the algebraic part of the proof. It shows that the 
$SU(N)$ algebra, under the rescaling (\ref{rescsu2gen}), does go to that of
$[SDiff(S^2)]$. Since the classical fields of an $SU(N)$ Yang-Mills theory can
be expanded in the basis of the matrices $T^{(N)}_{l,m}$, the relation
(\ref{limpois}) proves also the field theoretical result. 

\vskip 0.5cm

Here we want to argue that the equivalence between Yang-Mills
theories and field theories on surfaces is in fact valid to any order in
$1/N$. Coming back to the case of the sphere, because of the condition
(\ref{casimir}), we can parametrise the $T_i$'s in terms of two operators,
$z_1$ and $z_2$. As a first step we write:

\begin{equation}
\label{fuzsph}
T_{1}=cosz_1 ~(1-z_2^2)^{\frac{1}{2}},~~~~T_{2}=sinz_1 ~(1-z_2^2)^{\frac{1}{2}}, ~~~~T_{3}=z_2
\end{equation}

These relations should be viewed as defining the operators $z_1$ and
$z_2$. A similar parametrisation has been given by T. Holstein and H. Primakoff in terms of creation and annihilation operators \cite{hp}. At the limit of $N$ $\rightarrow$ $\infty$, they become the coordinates
$\phi$ and $cos\theta$ of a unit sphere. To leading order in  $1/N$, the commutation
relations (\ref{rescsu2}) induce the commutation relation (\ref{fuzsphcom}) between
the $z_i$'s: 

\begin{equation}
[z_1,z_2]=\frac{2i}{N}
\end{equation}

\noindent $i.e.$ the $z_i$'s satisfy a Heisenberg commutation relation with the unity
operator at the right-hand side. $1/N$ plays the role of $\hbar$. In higher
orders, however, the definitions (\ref{fuzsph}) must be corrected because the
operators $T_1$ and $T_2$ are no more hermitian. It turns out that  
a convenient choice is to use $T_+$ and $T_-$. We thus write:

\begin{equation}
\begin{split}
\label{fuzsphnew}
T_+ & =T_1+iT_2=e^{\frac{iz_1}{2}}~(1-z_2^2)^{\frac{1}{2}}~e^{
  \frac{iz_1}{2}}\\
T_- & =T_1-iT_2=e^{-\frac{iz_1}{2}}~(1-z_2^2)^{\frac{1}{2}}~e^{-
  \frac{iz_1}{2}}\\
T_3 & =z_2
\end{split}
\end{equation}

In order to avoid any misunderstanding, let me emphasise that (\ref{fuzsphnew}) and (\ref{fuzsph}) are not invertible as matrix relations and do not imply a representation of the $z$'s in terms of finite dimensional matrices. 
The claim instead is that the $SU(N)$ algebra can be expressed in two equivalent ways:
We can start from the non-commutative coordinates of the fuzzy sphere  $z_1$
and $z_2$ which are assumed to satisfy the quantum mechanical commutation
relations (\ref{fuzsphcom}). Through (\ref{fuzsphnew}) we define
three operators $T_1$, $T_2$ and $T_3$. We shall prove that they satisfy
exactly, without any higher order corrections, the $SU(2)$ relations
(\ref{rescsu2}) and (\ref{casimir}) and, consequently, they can be used as
basis for the entire $SU(N)$ algebra. 
The opposite is also true. The $SU(2)$ commutation relations (\ref{rescsu2}) imply the quantum mechanical
relation (\ref{fuzsphcom}). We can express the $SU(2)$ generators $T_i$
$i=1,2,3$, through (\ref{fuzsphnew}), in terms of two operators $z_i$ $i=1,2$. We can
again prove that they satisfy the Heisenberg algebra (\ref{fuzsphcom}) to all orders
in $1/N$. 

We start with the first part of the statement, which is
straightforward calculation. 

We assume (\ref{fuzsphcom}) and we want to compute the commutator of $T_+$ and
$T_-$ given by (\ref{fuzsphnew}).   

\begin{equation}
\begin{split}
\label{proof1}
[T_+,T_-] & =e^{\frac{iz_1}{2}}~(1-z_2^2)~e^{-
  \frac{iz_1}{2}}~-~e^{-\frac{iz_1}{2}}~(1-z_2^2)~e^{\frac{iz_1}{2}}\\
 & =\left(e^{\frac{iz_1}{2}}~z_2~e^{-
  \frac{iz_1}{2}}\right)^2~-~\left(e^{-\frac{iz_1}{2}}~z_2~e^{\frac{iz_1}{2}}\right)^2
\end{split}
\end{equation}

A useful form of the Cambell-Haussdorf relation for two operators $A$ and $B$
is:

\begin{equation}
\label{CH}
e^A B e^{-A}=B+[A,B]+ \frac{1}{2}[A,[A,B]]+...
\end{equation}

Applying (\ref{CH}) to (\ref{proof1}), we obtain:

\begin{equation}
\label{proof2}
[T_+,T_-]=\left(\frac{1}{N}+z_2 \right)^2~-~\left(\frac{1}{N}-z_2 \right)^2~=~\frac{4}{N}z_2
\end{equation}

Similarly, we check that the other two commutation relations of $SU(2)$ are
satisfied. We can also compute the Casimir operator $T^2$ and we find
the value $1-1/N^2$ of (\ref{casimir}).

We proceed now with the proof of the opposite statement, namely the equivalence
between the $SU(2)$ and the quantum mechanical commutation relations. The essence of the story is that any corrections
on the r.h.s. of (\ref{fuzsphcom}) will affect the $SU(2)$ commutation relations (\ref{rescsu2}). The
argument is inductive, order by order in $1/N$. 

Let us start with the first term and write the general form of (\ref{fuzsphcom}) as:

\begin{equation}
\label{;}
[z_1,z_2]=\frac{1}{N}t_1(z_1,z_2)+O(\frac{1}{N^2})
\end{equation}
with $t_1$ some function of the $z$'s. Using (\ref{;}) we compute the $1/N$ term in the commutator of two $T$'s given by
(\ref{fuzsphnew}). If we assume that they satisfy the $SU(2)$ commutation
relations  we determine
$t_1(z_1,z_2)$:

\begin{equation}
\label{;;;;;;}
t_1(z_1,z_2)=2i
\end{equation}

We can now go back to (\ref{;}) and determine the next term in the
expansion. We write:

\begin{equation}
\label{!}
[z_1,z_2]=\frac{2i}{N}+\frac{1}{N^2}t_2(z_1,z_2)+O(\frac{1}{N^3})
\end{equation}

We look now at the $T_i$'s given by (\ref{fuzsphnew}) and compute  the commutator between $T_+$ and $T_3$ to order
$1/N^2$. Imposing the absence of such terms, we get:

\begin{equation}
\label{!!!!}
t_2(z_1,z_2)=0
\end{equation}

It is now clear how the induction works: We assume the commutator

\begin{equation}
\label{!!!!!}
[z_1,z_2]=\frac{2i}{N}+\frac{1}{N^k}t_k(z_1,z_2)+O(\frac{1}{N^{k+1}})
\end{equation}
and set the coefficient of the corresponding term in the $SU(2)$ commutation
relation equal to zero. This gives again:

\begin{equation}
\label{**}
t_k(z_1,z_2)=0
\end{equation}

\vskip 1cm

The commutation relation (\ref{fuzsphcom}) is the main step of the argument. Any
function $f$ of the $SU(N)$ generators, in particular any polynomial of the Yang-Mills
fields and their space-time derivatives, can be rewritten, using 
 (\ref{fuzsphnew}), as a function of $z_1$ and $z_2$. Since they satisfy the
quantum mechanics commutation relations (\ref{fuzsphcom}), the usual proof of
Moyal \cite{moyal} goes through and the commutator of two such functions $f$ and $g$ will
have an expansion in powers of $1/N$ of the form:

\begin{equation}
\label{moyal}
[f,g] \sim \frac{1}{N} \{f(z_1,z_2),g(z_1,z_2)\} + \frac{1}{N^2}
\left(\{\frac{\partial f}{\partial z_1}, \frac{\partial
  g}{\partial z_2}\} + \{\frac{\partial f}{\partial z_2}, \frac{\partial
  g}{\partial z_1}\} \right) +...
\end{equation}

\noindent with the Poisson brackets defined the usual way: 

\begin{equation}
\label{poisson} 
\{f,g\} = \left(\frac{\partial f}{\partial z_1} \frac{\partial
  g}{\partial z_2} - \frac{\partial f}{\partial z_2} \frac{\partial
  g}{\partial z_1}\right)  
\end{equation}

The first term in this expansion is unambiguous but the coefficients of the
higher orders depend on the particular ordering convention one may adopt. For
example, in the symmetric ordering, only odd powers of $1/N$ appear. 

For the symmetric ordering, we can introduce, formally, a *-product through:

\begin{equation}
\label{star}
f(z)*g(z)=exp(\xi~ \epsilon _{ij}~\partial _z^i \partial _w^j)f(z)g(w)|_{w=z}
\end{equation}

\noindent with $z=(z_1,z_2)$ and $\xi=\frac{2i}{N}$. The $SU(N)$ commutators
in the Yang-Mills Lagrangian can now be replaced by the *-products on the
non-commutative surface. This equality will be exact at any given order in the
$1/N$ expansion. This completes the proof of our statement. 

\vskip 0.5cm

Before closing, a few remarks:

The proof has been given only for the case of the sphere. The extension to the
two-dimensional torus is straightforward, using the results of reference
\cite{zachos2}. In principle, however, such a formulation should be possible
for arbitrary genus surfaces, \cite{novikov}, \cite{jaffe}, although I do not
know of any explicit proof. 

It is straightforward to generalize these results and include matter fields,
provided they also belong to the adjoint representation of $SU(N)$. In
particular, the supersymmetric Yang-Mills theories have the same property. The
special case of ${\cal N}=4$ supersymmetry is of obvious interest because of
its conformal properties. In this theory the duality $g \rightarrow 1/g$ makes
us believe that 
 the two large $N$ limits, namely 't Hooft's and the one used here, are
 related.

 We believe that one could also
include fields belonging to the fundamental representation of the gauge
group. In 't Hooft's limit such matter multiplets are restricted to the
edges of the diagram, so we expect in our case the generalization to
involve open surfaces.

The equivalence between the original $d$-dimensional Yang-Mills theory and the
new one in $d+2$ dimensions holds at the classical level. For the new
formulation however, the ordinary perturbation series, even at the large $N$ limit, is divergent. The reason is that the quadratic
part of this action does not contain derivatives with respect to
$z_1$ or $z_2$. This is not surprising because these
divergences represent the factors of $N$ in the diagrams of the
original theory which have
not been absorbed in the redefinition of the coupling constant.
However, we expect a perturbation expansion around some appropriate
non-trivial classical solution to be meaningful and to contain
interesting information concerning the strong coupling limit of the
original theory. 

A final remark: Could one have anticipated the emergence of this action in the
$1/N$ expansion? It is clear that, starting from a set of $N$ fields
$\phi^i(x)$ $i=1,...,N$, we can
always replace $\phi^i(x)$, at the limit when $N$ goes to
infinity, with $\phi (\sigma, x)$ where $0 \leq \sigma \leq 2 \pi
$. In this case the sum over i will become an integral over
$\sigma $. However, for a general interacting field theory, the $\phi^4$ term  will no more be local in
$\sigma $. So, the only surprising feature  is that,
for a Yang-Mills theory, the resulting expression is local.

\end{document}